\documentclass[aps,reprint,prreview,superscriptaddress,twocolumn]{revtex4-2}
\usepackage{graphicx}
\usepackage{physics}
\usepackage{amsfonts}
\usepackage{geometry}
\usepackage{siunitx}
\usepackage[T1]{fontenc}
\usepackage{subfigure}
\usepackage[hidelinks]{hyperref}
\usepackage{braket}

\begin{document}
\title{Frequency-Axis Calibration via Autler-Townes Splitting in an Electromagnetically Induced Transparency Feature}

\author{Stone B. Oliver}
\email[]{stone.oliver@nist.gov}
\affiliation{National Institute of Standards and Technology, Boulder, Colorado 80305, USA}
\affiliation{Department of Physics, University of Colorado, Boulder, Colorado 80309, USA}

\author{Noah Schlossberger}    	
\affiliation{National Institute of Standards and Technology, Boulder, Colorado 80305, USA}

\author{Tate McDonald}
\affiliation{National Institute of Standards and Technology, Boulder, Colorado 80305, USA}
\affiliation{Department of Physics, University of Colorado, Boulder, Colorado 80309, USA}

\author{Christopher L. Holloway}
\affiliation{National Institute of Standards and Technology, Boulder, Colorado 80305, USA}

\author{Nikunjkumar Prajapati}    	
\email[]{nikunjkumar.prajapati@nist.gov}
\affiliation{National Institute of Standards and Technology, Boulder, Colorado 80305, USA}

\begin{abstract}
Rydberg atom-based electrometry is a pivotal technology that enables radio-frequency (rf) field measurements traceable to the International System of Units. This is typically accomplished by converting the direct rf power measurement of classical probes to a frequency measurement referenced to two well-defined spectral features, a commercial wavemeter, or narrow cavity lines produced via clock disciplined electro-optic modulators, although there exist cases where none of these are available. In this article, we demonstrate that a known rf frequency scale can be used to infer a laser's frequency scan using Autler-Townes (AT) splitting. In other words, we show that frequency-axis calibration can be achieved using the avoided-crossing behavior of near-resonant rf radiation. The relative separation between AT-split features is measured as a function of an applied rf field's power and detuning from atomic resonance. These peak separations are then fit to a simple, two-level model using a weighted nonlinear least-squares regression analysis that utilizes three fitting parameters, one of which is the desired calibration factor. While we do not benchmark this method against an absolute frequency reference, we do show that this method is (1) consistent with calibration via known fine-structure separation to within approximately 1.11\% or better, and (2) robust against substantial experimental deviation from the system's known rf power dependence.
\end{abstract}

\maketitle

\section{Introduction}
\label{IntroductionSection}
Electric field calibration (EFC) is critical to optimizing electromagnetic compatibility and plays a central role in accurate power calibration, but conventional EFC systems are constrained by measurement uncertainties no better than 5\% and an inability to link measurements directly to the International System of Units (SI) \cite{Simons_2024, Matloubi_1993, Hill_1990}. Such measurements are crucial in controlling the radiation emitted by or reflected from surfaces \cite{Paul_2023}, and are used in a variety of contexts, from radar technologies \cite{Freeman_2002} and channel sounding \cite{Mao_2024} to targeted tissue heating \cite{Vander_2006} and MRI exposure-risk assessment \cite{Hartwig_2022}.

Despite the relatively large uncertainties in EFC measurements, conventional calibration services are often time consuming and expensive, which has led to a growing interest in Rydberg atom-based EFC platforms. Rydberg atoms offer a single shot measurement of the electric field within a laser-atom interaction volume, and they are traceable to the SI system via time and Planck's constant \cite{Babusenan_2026, Schlossberger_2024_Gen, Anderson_2021, Simons_2018, Holloway_2014, Shaffer_2012}. Rydberg atom-based EFC systems have the potential to surpass current EFC capabilities by producing measurement results with smaller uncertainties and/or providing SI traceable EFC \cite{Holloway_2017-2}. 

Rydberg atom-based electric field sensors typically rely on electromagnetically induced transparency (EIT), a two-photon process that generates a narrow line measurement of a Rydberg state \cite{Finkelstein_2023, Derose_2023}. When subjected to an external radio-frequency (rf) field, the atomic response is governed by either Autler-Townes (AT) splitting or Stark shifting. The Stark shift dominates the off-resonant response, while AT-splitting occurs when the rf frequency lies near a Rydberg-Rydberg transition in the atom. In either case, the shift of the atomic lines can be used to determine the strength of the incident field \cite{Schlossberger_2024_Gen}. Hence, spectroscopically probing Rydberg atoms allows for calibrated electric field measurements. However, the laser frequency scan used to probe the Rydberg state must be calibrated to accurately measure these line shifts. Frequency-axis calibration is typically achieved using (1) two spectral features with known frequency separation, (2) a commercially available wavemeter, or (3) atomic-clock-disciplined electro-optic modulators which generate narrow sidebands for frequency reference. In some cases, additional features may not be available for reference [e.g., narrow frequency scan range requirements or systems where nearby (hyper) fine states are either unavailable or unresolvable], and the latter two options are both expensive and relatively bulky, making them less desirable for cost-constrained and/or field-deployable systems.

In this article, we provide a proof-of-principle demonstration of the frequency-axis calibration of a Rydberg atom system utilizing AT-splitting. As the rf field and coupling laser are detuned from resonance, the AT-split features display an avoided-crossing behavior that can be described using a simple, two-level model. Applying a weighted nonlinear least-squares regression (WNLSR) fit to the avoided-crossing curve yields a calibrated frequency axis. In essence, this method transfers the frequency scale of a known rf detuning onto a laser's frequency scan through use of the generalized Rabi frequency. We compare our calibration method against calibration via known fine-structure peak separation. We do not rigorously benchmark the result against an absolute frequency reference, although two-feature-based calibration is commonly accepted as a sufficiently robust method for many applications (see, e.g., \cite{Allinson_2024, Thakran_2024, su_2022, Prajapati_2021, Holloway_2018, Simons_2018}). The two methods are shown to agree to within approximately 1.11\% or better. We also demonstrate that calibration can still be achieved in cases where the experimental data deviates substantially from the model's known power dependence (e.g., calibration can be achieved in cases where there exists large uncertainty and/or error in the rf power incident on the atoms).

\section{Experiment} \label{experimental-setup_section}
\label{ExperimentalSetupSection}

\begin{figure}[!h]
\includegraphics[width=.999\linewidth]{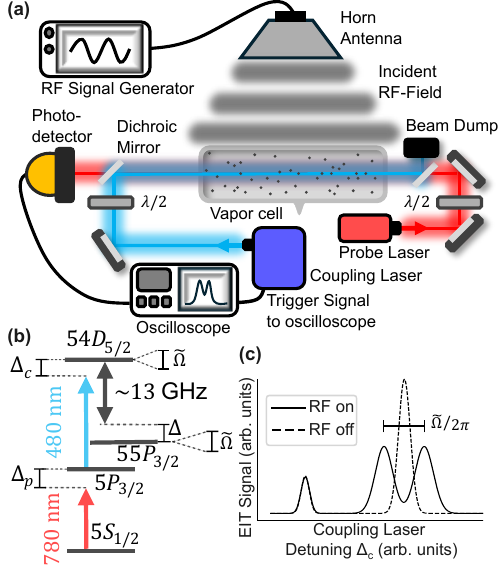}
\caption{\textbf{(a)} A schematic of the experimental setup. \textbf{(b)} The energy level structure used for this work. \textbf{(c)} Idealized EIT measurements either with (solid line) or without (dashed line) a resonant rf field applied. The symmetry of the AT-splitting is broken if the rf field is detuned from resonance. The smaller feature to the left is EIT resulting from nearby fine structure. This additional feature is commonly used to frequency calibrate the system.}
\label{Figure1}
\end{figure}

We deploy a standard two-photon EIT spectroscopy system, as shown in Fig. \ref{Figure1}(a). A $\SI{780}{\nano\meter}$ probe laser is tuned to the $D2$ transition for $^{85}$Rb, and a $\SI{480}{\nano\meter}$ coupling laser links the intermediate $5P_{3/2}$ state with the $54D_{5/2}$ state (see Fig. \ref{Figure1}(b)). The $1/e^{2}$ beam waist diameter for both beams was set to approximately $\SI{1.6}{\milli\meter}$. The probe and coupling laser powers are set to approximately $\SI{5}{\micro\watt}$ and $\SI{400}{\milli\watt}$, respectively. Both beams are linearly polarized along the same axis before being directed into a vapor cell containing room-temperature rubidium atoms. The beams are placed counter-propagating with respect to one another to partially cancel Doppler effects. EIT is observed by measuring the probe transmission through the vapor cell as the coupling laser frequency is scanned. Additionally, the frequency scan is made large enough to observe EIT resulting from both the $54D_{5/2}$ and the $54D_{3/2}$ states so that calibration results can be benchmarked against the two-feature approach. The frequency separation of these states was calculated using the Alkali-Rydberg calculator (ARC) in Python \cite{ARC_2017}. This calculation relies on data presented in Ref. \cite{Li_2003}; propagating uncertainties in the quantum defects reported there yields a result of 73.1791$\pm$0.0002 MHz.

Next, an rf horn antenna is placed such that the atoms are in the horn's Fraunhofer region (i.e., in the radiative far-field). The horn antenna's output is tuned to the $54D_{5/2}$ to $55P_{3/2}$ transition frequency, which was calculated to be 13.433374 GHz using ARC (uncertainty in this frequency is inconsequential for this work; see Sec. \ref{Calibration_Methodology}). Critically, this tone is produced using an rf generator referenced to a $\SI{10}{\mega\hertz}$ clock source. The rf polarization is also matched to that of the EIT beams to maximize the AT-splitting amplitude, and the vapor cell is insulated with rf-absorbing material to minimize stray fields within the laser-atom interaction volume. Fig. \ref{Figure1}(c) displays an idealized measurement output either with or without a resonant rf field applied.

The nominal power $P$ and nominal frequency detuning $\delta$ of the rf field are varied and the resulting AT-split EIT feature is recorded using an oscilloscope synchronized with the coupling laser's frequency scan. Here, 'nominal' refers to values selected by the experimenter, while 'actual' is used to describe the true values incident on the atoms. In this work, $P$ was swept from -23 dBm to -5 dBm in increments of 0.5 dBm. At each $P$, $\delta$ was swept from $\SI{-50 }{\mega\hertz}$ to $\SI{120}{\mega\hertz}$ in $\SI{5}{\mega\hertz}$ increments. The minimum power was selected based upon the lowest value at which AT-splitting could still be clearly observed at zero detuning. The minimum detuning was selected to ensure no interference between the Autler-Townes split features and the $54D_{3/2}$ feature. The maximum power and detuning values are constrained by (1) the signal-to-noise ratio and (2) the coupler scan range. The oscilloscope used to collect data was set to record an average of four traces. Since the calibration process relies on peak locations, but not peak widths, artificial broadening due to center frequency drift in the coupling laser scan during oscilloscope averaging is inconsequential for this work. The process was repeated five times, effectively yielding twenty total traces, the average of which is analyzed. Alongside each measurement, a reference measurement with no applied rf field is also recorded.

\section{Theory} 
\label{theory_section}
Here, we begin with a four-level Rydberg EIT system and show that AT-splitting in the EIT feature can be well-approximated using a two-level model. We then use the two-level model to develop an expression for the probe-transmission amplitude as a function of rf power, rf detuning, and coupling laser detuning. For this work, we use $\Omega$ to denote Rabi rates and $\Delta$ to denote detunings. The subscript $p$ refers to the ($\SI{780}{\nano\meter}$) probe field while the subscript $c$ refers to the ($\SI{480}{\nano\meter}$) coupling field. $\Omega$ and $\Delta$ without any subscript refer to the (actual) Rabi rate and (actual) detuning of the rf field, respectively. Generally, the Rabi rate resulting from incident electric field $\vec{E}_{i}$ is described by $\Omega_{i} = \vec{d} \cdot \vec{E_{i}}/\hbar$ where $\vec{d}$ is the dipole moment of the transition induced by $\vec{E_{i}}$ and $\hbar$ is the reduced Planck constant \cite{Berman_2011}.

Working in the interaction picture, assuming a weak, resonant probe ($\Delta_{p}=0$), and assuming real Rabi rates for all fields, the semi-classical dressed-state Hamiltonian for the four-level system shown in Fig. \ref{Figure1}(b) is
\begin{equation}
\begin{split}
H = - \frac{\hbar}{2} & \big( 2\Delta_{c} \ket{r_{1}} \bra{r_{1}} + 2(\Delta_{c} - \Delta) \ket{r_{2}} \bra{r_{2}} \\
& + \Omega_{p} \left( \ket{g} \bra{i} + \ket{i} \bra{g} \right) \\
& + \Omega_{c} \left( \ket{i} \bra{r_{1}} + \ket{r_{1}} \bra{i} \right) \\
 & + \Omega \left( \ket{r_{1}} \bra{r_{2}} + \ket{r_{2}} \bra{r_{1}} \right)   \big),
\end{split}
\end{equation}
where, for this work, $\ket{g}$, $\ket{i}$, $\ket{r_{1}}$, and $\ket{r_{2}}$ correspond to the $5S_{1/2}$, $5P_{3/2}$, $54D_{5/2}$, and $55P_{3/2}$ states, respectively \cite{Manchaiah_2022}. In matrix form,
\begin{equation} \label{H_Matrix_Eqn}
\begin{split}
&H = -\frac{\hbar}{2}\begin{pmatrix}
    0 & \Omega_{p} & 0 & 0 \\
    \Omega_{p} &  0 & \Omega_{c} & 0  \\
    0 & \Omega_{c} &  2\Delta_{c} & \Omega  \\
    0 & 0 & \Omega & 2(\Delta_{c} - \Delta) 
\end{pmatrix}_{.}
\end{split}
\end{equation}
With no rf field applied, the second Rydberg state $\ket{r_{2}}$ may be ignored. The atoms interacting with the EIT beams enter a dark state where probe transmission is induced by coherent interaction with $\ket{r_{1}}$ despite zero intermediate population. Once the rf field is switched on, the Rydberg states $\ket{r_{1}}$ and $\ket{r_{2}}$ become coupled; the resulting two-level sub-system has dressed states constructed from linear combinations of the Rydberg states. Thus, the resulting probe transmission is effectively a projective measurement of the dressed states onto $\ket{r_{1}}$. Since the rf field only couples the Rydberg states, we need only consider the lower-right 2$\times$2 sub-block in our matrix expression for the Hamiltonian, denoted by $H_{2}$, with the understanding that the dynamics of this two-level sub-system will be encoded in the probe coherence (i.e., in the probe transmission). Shifting the zero-point energy by $+ \hbar \Delta_{c}$,
\begin{equation}
H_{2} = - \frac{\hbar}{2} \begin{pmatrix} 0 & \Omega \\ \Omega  & -2 \Delta
    \end{pmatrix}_{.}
\end{equation}
This is the standard semi-classical two-level dressed-state Hamiltonian \cite{Berman_2011}. The eigenenergies of this Hamiltonian are
\begin{equation} \label{eigenenergies_eqn}
E_{\pm} =  \frac{\hbar \Delta}{2} \pm \frac{\hbar \sqrt{\Delta^{2} + \Omega^{2}}}{2},
\end{equation}
meaning that the frequency separation between the dressed states is given by the generalized Rabi frequency,
\begin{equation} \label{GeneralizedRabiEqn}
\tilde{\Omega} = \sqrt{\Delta^{2} + \Omega^{2}}.
\end{equation}
The dressed (eigen) states are
\begin{equation}
\ket{\pm} = \frac{\frac{\Delta \mp \sqrt{\Delta^{2} + \Omega^{2}}}{\Omega} \ket{r_{1}} + \ket{r_{2}}}{\sqrt{ \left| \frac{\Delta \mp \sqrt{\Delta^{2} + \Omega^{2}}}{\Omega} \right|^{2} + 1}}.
\end{equation}
Since the probe transmission amounts to projection onto $\ket{r_{1}}$, the relative amplitudes of these dressed states are given by $A_{\pm} \equiv \left|\braket{r_{1}|\pm}\right|^{2}$, which can be written as
\begin{equation} \label{Amplitude_eqn}
A_{\pm}  = \left| \frac{\frac{\Delta \mp \sqrt{\Delta^{2} + \Omega^{2}}}{\Omega}}{\sqrt{  \left| \frac{\Delta \mp \sqrt{\Delta^{2} + \Omega^{2}}}{\Omega} \right|^{2} +1}}  \right|^{2}.
\end{equation}

\begin{figure} [!h]
\includegraphics[width=1.0\linewidth]{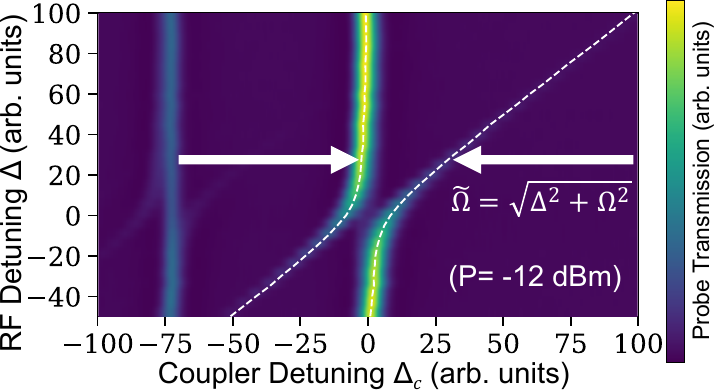}
\caption{Above displays an avoided-crossing in the AT-splitting of an EIT feature as $\Delta_{c}$ and $\Delta$ are varied. Such behavior is well-described by Eq. \ref{probe_amplitude_eqn}. Data was collected using $P=-12$ dBm, and the unperturbed feature is EIT due to the $54D_{3/2}$ state.}
\label{Avoided_Crossing_Figure}
\end{figure}

As shown in Fig. \ref{Avoided_Crossing_Figure}, varying $\Delta_{c}$ and $\Delta$ while holding $\Omega$ fixed yields an avoided-crossing behavior. The eigenenergies in expression \ref{eigenenergies_eqn} provide the position of these features relative to zero coupler detuning, and expression \ref{Amplitude_eqn} provides their relative amplitudes \cite{Schlossberger_2025}. Since the experiment is performed using room-temperature atoms in a cell with no buffer gas, these AT-split lines are also expected to exhibit Gaussian profiles owing to the dominance of inhomogeneous Doppler broadening over other (homogeneous) broadening mechanisms \cite{Finkelstein_2023}. Thus, the probe transmission amplitude can be approximated as
\begin{equation} \label{probe_amplitude_eqn}
\begin{split}
A_{p}(\Delta_{c}, \Delta, \Omega) = & A_{+}e^{-(\hbar \Delta_{c}-E_{+})^{2}/2 \sigma_{+}^{2}} \\
& + A_{-}e^{-(\hbar \Delta_{c}-E_{-})^{2}/2 \sigma_{-}^{2}}
\end{split}
\end{equation}
where $\sigma_{\pm}$ refers to the Gaussian root-mean-square (RMS) width of the profiles.

\section{Calibration Methodology}
\label{Calibration_Methodology}
While Eq. \ref{probe_amplitude_eqn} provides probe-transmission amplitude as a function of $\Delta_{c}$, $\Delta$, and $\Omega$, probe transmission is typically recorded using an oscilloscope that records voltage as a function of time $t$. Furthermore, only the nominal rf power and detuning, $P$ and $\delta$, are known. Assuming an idealized, linear sweep of $\Delta_{c}$ as the oscilloscope collects data, frequency-axis calibration amounts to determining the amount by which $\Delta_{c}$ changes over any given duration of time.

In principle, the rf field can be fixed perfectly on resonance, although in practice, there exists uncertainty in the rf source's frequency output, and atomic lines are shifted as a result of not only external stray fields but also fields generated by charges residing within the vapor cell used for the experiment \cite{Patrick_2025}. To account for any unknown offset in resonant frequency, we decompose the detuning as $\Delta = 2 \pi(\delta + \epsilon)$, where $\epsilon$ is assumed constant. Additionally, the rf power incident on the atoms differs from the nominal power supplied by the rf source as a result of attenuation in the rf cables leading to the horn antenna, the gain of the horn antenna and its distance from the atoms, and scattering and reflections in the environment. Since the quantities $P$ and $\Omega^{2}$ are both proportional to the square of the rf field, we suppose $\Omega^{2}= 4 \pi^{2} \alpha P$ where $\alpha$ is an unknown attenuation constant. We also introduce a calibration factor with units of time squared, denoted by $C$, which converts the passage of time to a shift in coupler detuning $\Delta_{c}$.

 \begin{figure} [!h]
\includegraphics[width=0.9\linewidth]{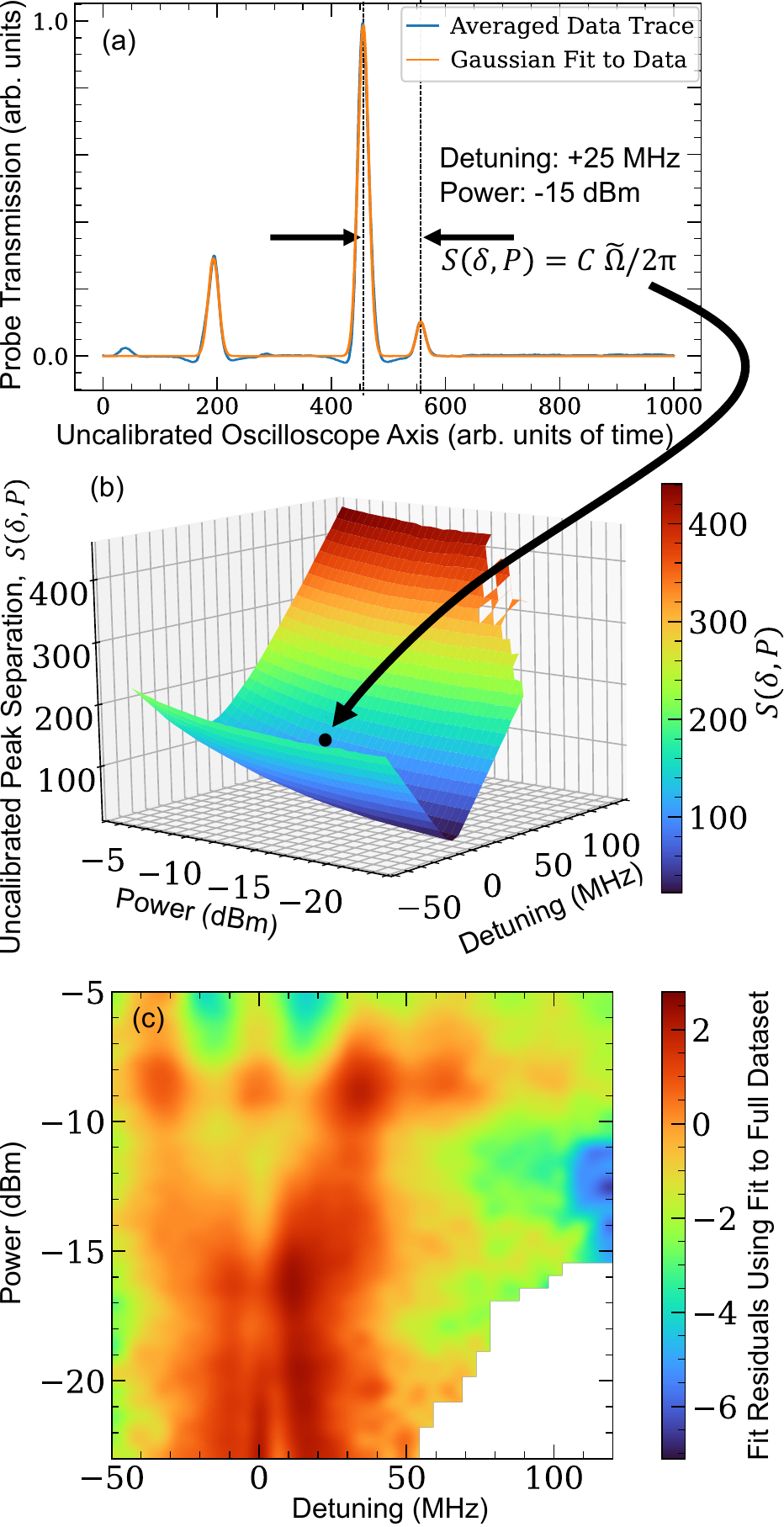}
\caption{\textbf{(a)} Data collected at $\delta = \SI{25}{\mega\hertz}$ and $P = -15$ dBm and the corresponding fit of Gaussian curves to the dataset (see Eq. \ref{probe_amplitude_eqn}). The uncalibrated AT-split peak separation $S(\delta, P)$ is calculated using the peak locations of the fitted curve. \textbf{(b)} A plot of $S(\delta, P)$. The black dot corresponds to the value of $S(\delta, P)$ extracted from the data shown in (a). This surface is then fit to Eq. \ref{GeneralizedRabiEqn_FitFunc} using WNLSR with $C$, $\epsilon$, and $\alpha$ as fitting parameters. \textbf{(c)} The residuals resulting from fitting the data shown in (b) to Eq. \ref{GeneralizedRabiEqn_FitFunc}.}
\label{Gauss_fit_fig}
\end{figure}

The calibration process begins by fitting each averaged oscilloscope trace to Eq. \ref{probe_amplitude_eqn}, where the amplitudes, center positions, widths, and a global DC offset are used as fitting parameters. Since our data includes AT-split peaks and an additional fine-structure peak, we fit three Gaussians to each trace; a sample trace and its corresponding fit are shown in Fig. \ref{Gauss_fit_fig}(a). Using these results, the center positions of each of the AT-split features are extracted to determine the uncalibrated peak separation as a function of $\delta$ and $P$, denoted by $S(\delta, P)$ (see Fig. \ref{Gauss_fit_fig}(b)). When $P$ is small but $\delta$ is large, the smaller AT-split feature may fall below the noise floor, in which case the Gaussian curve-fitting process fails and no peak separation is recorded. Rewriting Eq. \ref{GeneralizedRabiEqn} in terms of $S(\delta, P)$,
\begin{equation} \label{GeneralizedRabiEqn_FitFunc}
\begin{split}
\frac{\tilde{\Omega}}{2 \pi} = \frac{S(\delta, P)}{C} & = \sqrt{(\delta + \epsilon)^{2} + \alpha P} \\
\implies S(\delta, P) & = C\sqrt{(\delta + \epsilon)^{2} + \alpha P}, \\
\end{split}
\end{equation}
where $\tilde{\Omega}/ 2 \pi$ is the AT-split peak separation in units of frequency. Although $P$ was swept using logarithmic units, note that here, $P$ must be expressed using linear units (e.g., mW). Using $C$, $\epsilon$, and $\alpha$ as fitting parameters, the uncalibrated peak separations are fit to Eq. \ref{GeneralizedRabiEqn_FitFunc} using a WNLSR analysis, where each separation is weighted by an estimate of its uncertainty (see Sec. \ref{Results_Error_Analysis}). The resulting value of $C$ then converts the oscilloscope's time axis to frequency detuning, where the zero-detuning point is taken to be the center position of the EIT feature observed when no rf field is present.

\section{Results and Discussion}
\label{Results_Section}

\subsection{Fine-Structure Calibration and Calibration Variation Noise}
\label{Results_FSC_Section}

\begin{figure} [h!]
\includegraphics[width=1.0\linewidth]{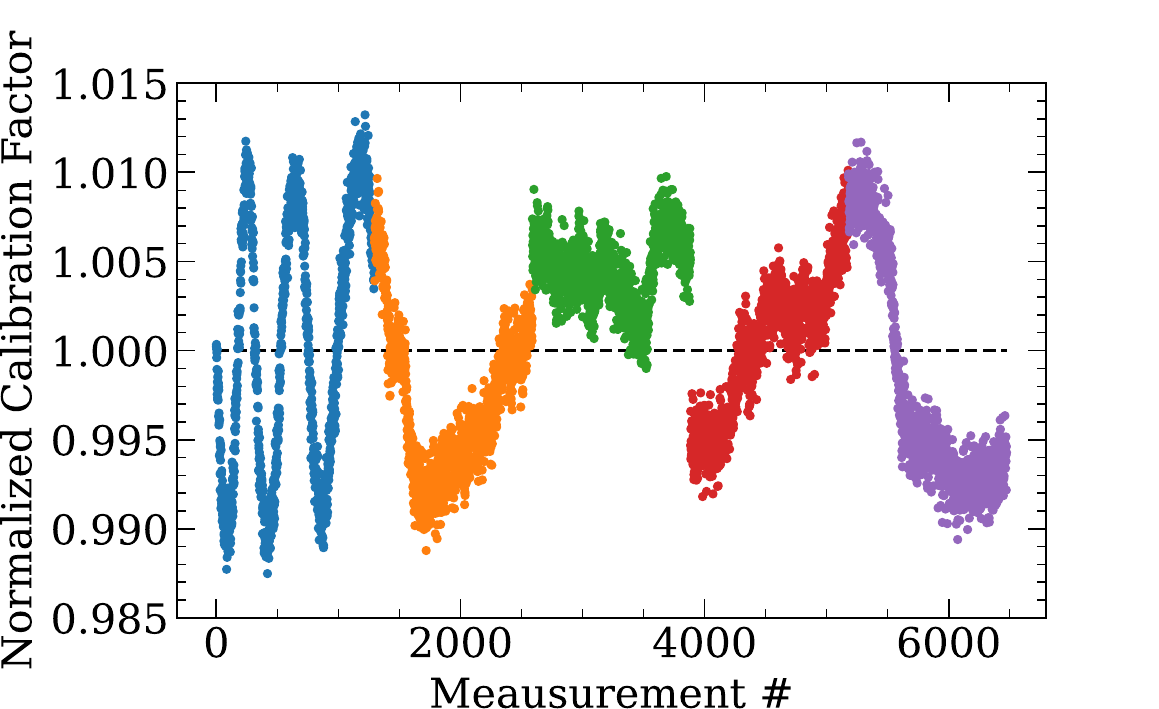}
\caption{The figure above plots fine-structure calibration results, $C_{fs}$ across measurement runs. Each color in the scatterplot data represents a different measurement run. Each run took approximately 45 minutes to complete. Results have been normalized to the mean.}
\label{FSC_Variation_Figure}
\end{figure}

This work assumes an idealized, linear frequency scan, although both calibration methods are susceptible to error arising from variation in the scan. For example, thermal fluctuations within the laser diode can influence the exact frequency produced by the applied modulation \cite{ito_2003, Zyoud_2020, Nadezhdinskii_1995}. Frequency scan error can manifest as center frequency drifts, scan nonlinearity, and/or variation in the scan range. Sufficiently slow drift in the scan's center frequency is accounted for using the reference traces recorded alongside each measurement; each reference trace is cross-correlated with a global reference trace, and the resulting shift is applied to its corresponding data trace. This process cannot account for drifts in center frequency that occur faster than the time required to collect a measurement and its corresponding reference trace (for this work, about two seconds), as such drift induces a mismatch in the center frequency of a measurement and its corresponding reference trace. Center frequency drifts occurring on timescales faster than the scan-rate of the frequency scan manifest as nonlinearity. Assuming the system is capable of producing a sufficiently linear scan modulation at the given frequency, the use of a faster scan repetition rate ($\SI{40}{\hertz}$ for this work) serves to reduce such nonlinearity.

Fine-structure calibration across all $N=6475$ reference traces yielded a mean calibration result of approximately $\bar{C}_{fs}=3.639$ (with appropriate units that convert MHz to oscilloscope pixels) and a percent standard deviation of $100(\sigma_{fs}/ \bar{C}_{fs}) = 0.57\%$. Error contributions from uncertainty in the frequency separation calculated using ARC and the Gaussian curve-fitting process were negligible, indicating that run-to-run variation is dominated by (1) frequency scan noise and/or (2) stray-field-induced lineshifts which vary over time. We refer to these effects collectively as calibration variation noise (CVN). Fig. \ref{FSC_Variation_Figure} plots the observed variation. Given percent-level variation in the frequency scan over the duration of the experiment, calibrating via AT-splitting is expected to exhibit comparable (or larger) uncertainty.

\begin{figure*}
\includegraphics[width=0.8\linewidth]{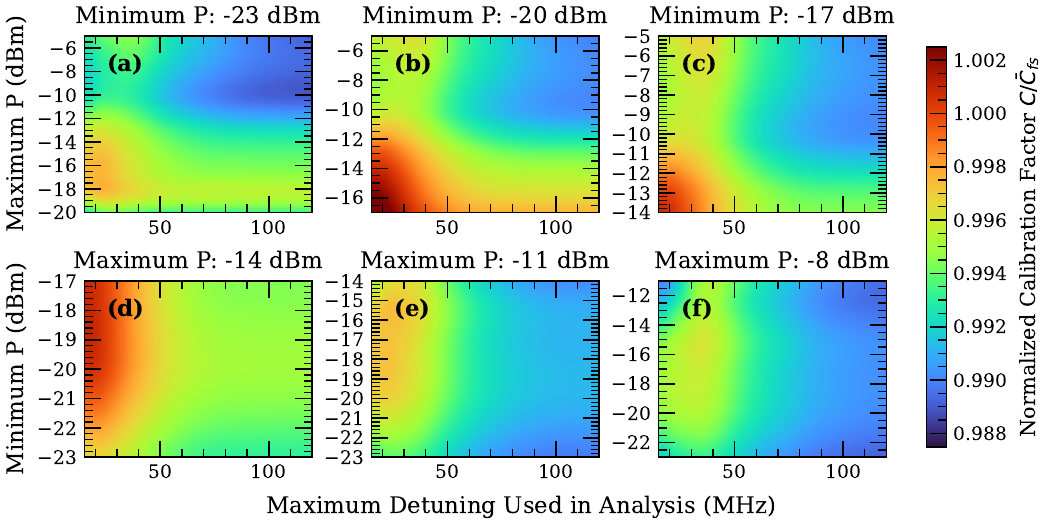}
\caption{Above displays values of $C / \bar{C}_{fs}$. Values correspond to data over different ranges of $P$ and $\delta$ used in the fitting process. For all data presented above, the minimum $\delta$ used for the analysis was fixed at $\SI{-50}{\mega\hertz}$. \textbf{(a)} $C / \bar{C}_{fs}$ as a function of the maximum values of $P$ and $\delta$ used in the analysis with the minimum value of $P$ fixed at -23 dBm. The top right-most value in this sub-plot corresponds to the result obtained using all available data (about 0.989). \textbf{(b)} The same plot as in (a) except the minimum value of $P$ has been fixed at -20 dBm. \textbf{(c)} The same plot as in (a) except the minimum value of $P$ has been fixed at -17 dBm. \textbf{(d)} $C / \bar{C}_{fs}$ as a function of the minimum value of $P$ and maximum value of $\delta$ used in the analysis with the maximum value of $P$ fixed at -14 dBm. \textbf{(e)} The same plot as in (d) except the maximum value of $P$ has been fixed at -11 dBm. \textbf{(f)} The same plot as in (d) except the maximum value of $P$ has been fixed at -8 dBm.}
\label{Calibration_Results_Fig}
\end{figure*}

\subsection{Calibration via AT-Splitting}
\label{Results_Autler_Townes_Calibration}


\begin{table}
\includegraphics[width=1.0\linewidth]{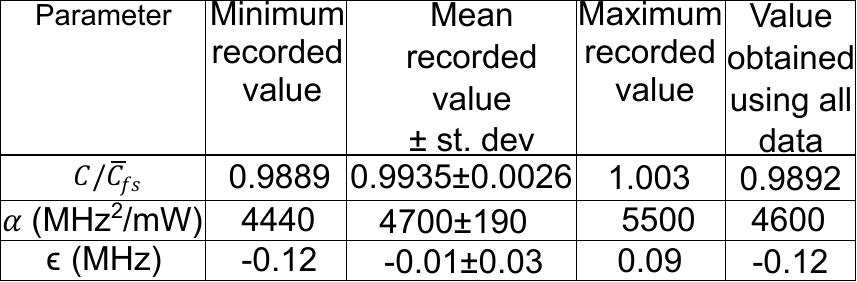}
\caption{The table above displays the minimum, mean $\pm$ standard deviation, and maximum values of $C/\bar{C}_{fs}$, $\alpha$, and $\epsilon$ found using data limited to various subsets of $\delta$ and $P$ to calibrate the system (see Fig. \ref{Calibration_Results_Fig}). It also reports the results of analyzing all available data.}
\label{Ref_Table} 
\end{table}

Fig. \ref{Calibration_Results_Fig} displays the AT-splitting-based calibration result normalized to the mean fine-structure result, $C / \bar{C}_{fs}$, as a function of the range of rf powers and detunings utilized for analysis. We calculate statistics on the result under the assumption of no preferred power or detuning ranges. Over all analyses, the AT-splitting calibration process was found to agree with fine-structure calibration to within approximately 1.11\% or better. Table \ref{Ref_Table} provides the ranges of $C/ \bar{C}_{fs}$, $\alpha$, and $\epsilon$ determined from these analyses. We also calculate the percent standard error associated with every determination of $C/ \bar{C}_{fs}$. Values were calculated by propagating uncertainty in (1) the fitting process and (2) the fine-structure calibration result. Since CVN induces variation in $\bar{C}_{fs}$ itself rather than contributing to measurement error, we treat $\bar{C}_{fs}$ as a true mean rather than a statistical one. As such, we propagate the variance of $C_{fs}$ rather than the variance of its mean (i.e., we propagate error using $\sigma_{fs}^{2}$, not $\sigma_{fs}^{2}/N$). The mean percent uncertainty was found to be 0.57\% with negligible standard deviation. Assuming uncertainty in $C$ from the WNLSR fit is well-estimated (read on), the percent standard error matching the percent standard deviation in $\bar{C}_{fs}$ is clear evidence of CVN dominating over other sources of variation. For comparison, the mean percent uncertainty obtained with CVN excluded was calculated to be only 0.026\%, although we show in section \ref{Results_Error_Analysis} below that a more realistic estimation of the total uncertainty is approximately 0.59\%, owing to underestimated uncertainty in the fitting process resulting from systematic errors. While CVN and fitting uncertainty provide a plausible explanation for the variation observed in Fig. \ref{Calibration_Results_Fig}, they fail to adequately explain why the mean value of $C / \bar{C}_{fs}$ lies approximately 0.65\% below unity.

\subsection{Error Analysis and Model Robustness Against Deviation From the \\ Idealized Power Response}
\label{Results_Error_Analysis}


Error induced by CVN, as well as error resulting from the Gaussian curve-fitting process were used to weight each point in the WNLSR fit (we use $\sigma = \sqrt{Cov(i,i) + Cov(j,j) - 2Cov(i,j)}$ to calculate the uncertainty in each point due to curve-fitting, where $Cov(i,j)$ is the $(i,j)$ component of the covariance matrix resulting from the fit and $i$ and $j$ denote the indices of that matrix corresponding to the peak locations of the corresponding oscilloscope trace). While CVN dominates the variation in $C / \bar{C}_{fs}$, uncertainty in the Gaussian curve-fitting process was found to dominate uncertainty in $S( \delta, P)$. The residual errors shown in Fig. \ref{Gauss_fit_fig}(c) were also found to yield a reduced chi-square value of $\chi^{2}_{R} \approx 43$ (the number of degrees of freedom was estimated as 1205 points minus three fit parameters), indicating a poor choice in model and/or severely underestimated uncertainties provided to the fitting model. While a small subset of points was found to inflate $\chi^{2}_{R}$ by roughly a factor of two, obvious systematic behavior in Fig. \ref{Gauss_fit_fig}(c) and the percent-level uncertainty observed in $S(\delta, P)$ lend credence to poor model choice being the primary source of $\chi^{2}_{R}$'s large magnitude.

\begin{figure} [h!]
\includegraphics[width=1.0\linewidth]{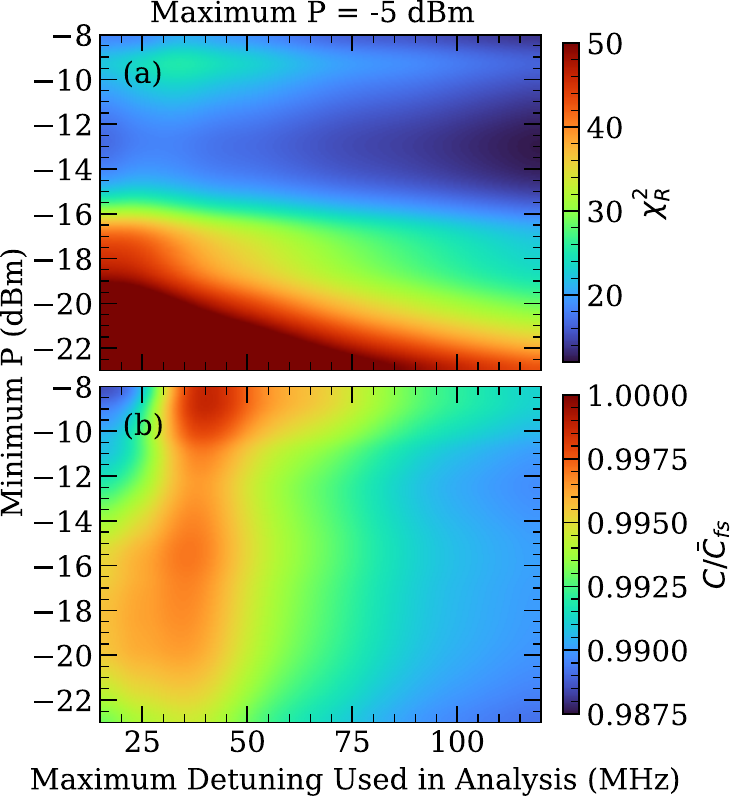}
\caption{(a) shows the quantity $\chi^{2}_{R}$ as a function of the minimum power and maximum detuning considered for analysis. (b) displays the corresponding value of $C / \bar{C}_{fs}$ for these analyses.}
\label{Fit_Quality_Analysis_Figure}
\end{figure}

It is interesting that this model's determination of $C$ is stable to within variation attributable to CVN across a wide range of rf power and detuning ranges, especially in light of the fact that the ranges of rf power and detuning selected for this work were limited by the experimental setup rather than guided by an optimization process. To investigate this effect, Fig. \ref{Fit_Quality_Analysis_Figure}(a) displays $\chi^{2}_{R}$ as a function of minimum power and maximum detuning considered in the analysis, and Fig. \ref{Fit_Quality_Analysis_Figure}(b) displays the value of $C/ \bar{C}_{fs}$ extracted from the corresponding analysis. While $\chi^{2}_{R}$ reduces in magnitude by roughly a factor of four as the minimum considered power is increased at any given detuning range, $C$ exhibits sub-percent variation (also note identical trends in Figs. \ref{Calibration_Results_Fig}(d) - \ref{Calibration_Results_Fig}(f)). This demonstrates empirically that there exists distinctly larger deviation from the prescribed model at lower powers that does not substantially impact the model's determination of $C$. This also directly implies that uncertainty in $C$ resulting from the WNLSR fit is likely underestimated.

Despite poor model fit, $C$ is thought to be robustly determined because it depends primarily on variation in $\delta$ rather than $P$ in the fitting process. Expanding Eq. \ref{GeneralizedRabiEqn_FitFunc},

\begin{equation}
\begin{split}
S( \delta, P) & = C \sqrt{(\delta + \epsilon)^{2} + \alpha P} \\
& = \sqrt{C^{2} \delta^{2} + 2C^{2} \epsilon \delta + C^{2} \epsilon^{2} + C^{2} \alpha P }.
\end{split}
\end{equation}

Clearly, the magnitude of the fitted peak separation in the limit of small $\delta$ and $P$ is set by $C\epsilon$, while $C^{2}$ and $2C^{2}\epsilon$ are extracted using variation in $S(\delta,P)$ as a function of $\delta$. The model presumably uses these three terms to disambiguate $C$ from $\epsilon$. This notion is well-supported by the Pearson correlation coefficient relating $C$ and $\epsilon$, denoted by $p_{C \epsilon}$, which was observed to be highly sensitive to the range of $\delta$ used for calibration; as the range of $\delta$ values was increased, $p_{C \epsilon}$ was observed to drop from upwards of 0.7 (high correlation) to below 0.2 (low correlation). On the other hand, $p_{C \alpha}$ was found to range from approximately -0.92 to -0.82 over all analyses. This strong correlation is consistent with the fact that $C$ and $\alpha$ become fully degenerate in the zero-detuning limit. High degeneracy between two parameters often yields larger uncertainty because nearly identical fit results are obtainable through simultaneous adjustments of both parameters.

Combined with the small uncertainties observed when CVN is excluded from the analysis (i.e., combined with the model's small and presumably underestimated confidence bounds on $C$), these observations imply that the model uses variation in $P$ to obtain $C^{2}\alpha$, but effectively optimizes $C$ independent of this parameter, in which case $\alpha$ is adjusted to ensure $C^{2} \alpha$ is consistent with the effectively predetermined value of $C$. In this situation, the degeneracy between $C$ and $\alpha$ is rendered inconsequential to calibration. However, this does not negate any error in the fit induced by unmodeled variation in either $\alpha$ or $P$ (e.g., frequency-dependent reflections, rf cable attenuation, or error in the rf power source's power output over the duration of the experiment). Hence, systematic error due to such variation provides a plausible explanation for why a poor fit to peak separation data has no obvious impact on $C$.

To test this hypothesis, uncalibrated peak separation data was simulated using
\begin{equation}
\label{nonlinear_power_eqn}
S(\delta, P) = C \sqrt{ (\delta + \epsilon)^{2} + \alpha P^{n} }
\end{equation}
with $C = \bar{C}_{fs}$, $\epsilon = 0$ MHz, and $\alpha  = 4700$ MHz$^{2}$/mW. By choosing $n \neq 1$, peak separations are systematically shifted from the values expected under Eq. \ref{GeneralizedRabiEqn_FitFunc}. To estimate a reasonable range of $n$ values, we introduce $n$ as an additional fitting parameter and fit the experimental data to Eq. \ref{nonlinear_power_eqn} in place of Eq. \ref{GeneralizedRabiEqn_FitFunc}. We repeat all analyses presented in Fig. \ref{Calibration_Results_Fig} in this fashion. The $\chi^{2}_{R}$ values resulting from these fits were still substantially larger than unity but were notably reduced compared to their original counterparts, indicating that the poor fit to the original model is at least partially a result of a breakdown of the assumption that $\Omega^{2} = 4 \pi^{2} \alpha P$ due to systematic experimental error(s). Over all analyses, $n$ was found to be 0.93 with a standard deviation of 0.04. As such, we chose to simulate data with $n$ ranging from 0.89 to 0.97 and compare the value of $C$ used to generate the simulated data against the result obtained by fitting that data to Eq. \ref{GeneralizedRabiEqn_FitFunc}. We take the simulated data to have identical uncertainty to the measured data to more closely replicate realistic experimental conditions.

\begin{figure} [h!]
\includegraphics[width=1.0\linewidth]{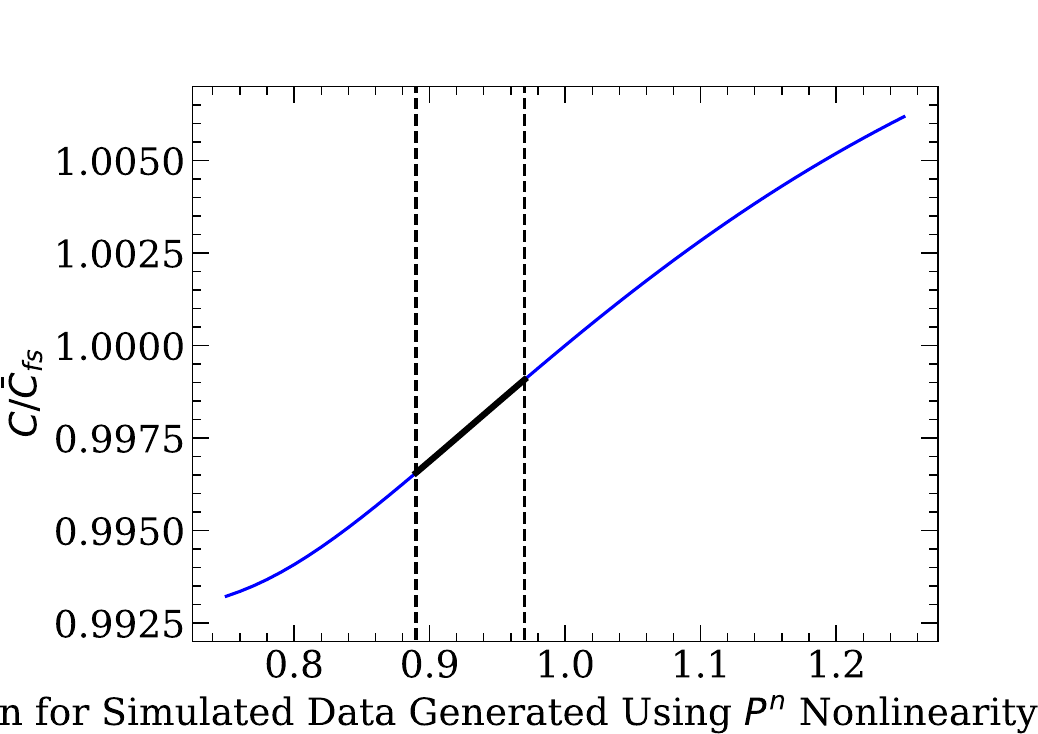}
\caption{Above plots $C/ \bar{C}_{fs}$ determined by fitting simulated peak separation data with $P^{n}$ dependence. The black line between the dashes corresponds to the range of $n$ values noted in the text. We take the range of $C/ \bar{C}_{fs}$ observed between the dashed lines to be a reasonable estimate of uncertainty in $C$ due to the fitting process.}
\label{P_nonlinearity_fig}
\end{figure}

Uncertainty in the determination of $C$ using simulated data remained on the order of 0.01\%, although $C$ was found to display substantially larger, highly linear variation as a function of $n$ over the relevant range (see Fig. \ref{P_nonlinearity_fig}). Since (1) the trend is linear, and (2) the minimum and maximum values of $n$ correspond to one standard deviation from the mean observed in the experimental data, half the observed range for $C / \bar{C}_{fs}$ corresponds to a reasonable estimate of uncertainty in the fitting process driven by experimental error/variation in either $\alpha$ and/or $P$. Hence, for the data shown in Fig. \ref{Calibration_Results_Fig}, we estimate a percent uncertainty due to the fitting process of approximately 0.13\%. Adding this error in quadrature with the CVN uncertainty of 0.57\% yields a total percent uncertainty of approximately 0.59\%.

This modest adjustment to the total uncertainty provides a realistic estimate of the expected variation in $C$ but still fails to address the observed 0.65\% offset of the mean calibration result from unity. However, it is noteworthy that the mean value of $C / \bar{C}_{fs}$ over $n \in (0.89, 0.97)$ in Fig. \ref{P_nonlinearity_fig} is observed to lie approximately 0.22\% below unity, indicating that $C$ may be systematically underestimated in our system as a result of $P^{n}$ dependence where $n < 1$. As such, effects which alter $\alpha$ or $P$ over the experiment, including systematic error in the rf generator's power output, power dependent rf cable attenuation, and frequency/power dependent reflections provide plausible explanations for why the AT-splitting-based calibration results are generally smaller than $\bar{C}_{fs}$, although the data does not fully support this hypothesis. Shifting the mean result up by 0.22\% still leaves the mean AT-splitting calibration result 0.43\% below unity, meaning the majority of the systematic offset remains unaccounted for.

We do not further address the observed underestimation of $C$. Beyond the sources of error discussed above, other effects that may influence the calibration process include data resolution limitations (which contribute to uncertainty in the Gaussian curve-fitting process) and error resulting from extraneous features inducing systematic offsets in the positions, amplitudes, and/or widths of the features of interest. For instance, magnetic sub-levels can produce extraneous spectral features when subjected to imperfect rf polarization (i.e., when subjected to elliptical polarization) \cite{Schlossberger_2026}.

\section{Conclusion}

In this work, we have provided a proof-of-principle demonstration of frequency-axis calibration using the avoided-crossing behavior of AT-splitting. The uncalibrated separation between AT-split peaks can be fit directly to a two-level model where frequency detuning offset $\epsilon$, power attenuation factor $\alpha$, and calibration factor $C$ are used as fit parameters, thus circumventing the fact that $\epsilon$ and $\alpha$ are unknown. In situations where the true incident detuning and power are accurately known, a fit could be applied with these values fixed (or highly constrained), thereby reducing calibration factor uncertainty. Applying the calibration method developed here, it is feasible to frequency calibrate systems without the use of a narrow-line cavity or an additional reference feature, a capability which is crucial for the development of budget and/or footprint constrained Rydberg sensors (e.g., field-deployable sensors).

The value of $C$ extracted from this method was found to be consistent with the result obtained using fine-structure separation to within approximately 1.11\% or better. Variation in $C$ as the ranges of rf powers and detunings considered in the analysis were varied was found to be dominated by CVN; the mean percent uncertainty in $C$ was estimated to be approximately 0.59\%. While the observed variation in $C$ is well-understood, effects which yield systematic underestimation of $C$ are not. However, experimental deviations from the prescribed rf power response alone cannot account for the observed offset. This was demonstrated by simulating peak separation data with $P^{n}$ dependence and observing the quality of the calibration result as a function of $n$. This analysis demonstrated systematic underestimation of WNLSR fit uncertainties and estimated the true uncertainty to be approximately 0.13\%, but it also showed that this effect could plausibly account for no more than roughly a 0.22\% shift in $C/ \bar{C}_{fs}$ from unity, leaving the majority of the observed offset unaccounted for. 

While the source of systematic underestimation remains unknown, this analysis demonstrates calibration via AT-splitting to be robust against substantial experimental departure from the idealized rf power response (i.e., systematic departure from the equation $\Omega^{2} = 4 \pi^{2}\alpha P$ due to experimental error). This results from the fact that $C$ is determined primarily through the recorded variation in peak separation as a function of $\delta$ (not $P$), which effectively negates deleterious effects resulting from (1) poor model fit as a function of $P$, and (2) large $C$-$\alpha$ degeneracy, both of which were observed in our analysis. This claim was supported empirically by observing sub-percent variation in $C$ across analyses of varying power range, whereas $\chi^{2}_{R}$ exhibited roughly a factor four shift in magnitude. This result also underscores the fact that this calibration method ultimately functions by transferring the known frequency scale of the clock source to the frequency scan axis through AT-splitting. This resilience will undoubtedly prove invaluable to field-deployable Rydberg-atom-based electric field sensors operating in noisy environments.

Future work may include analysis of the systematic errors left unaddressed in this work and the construction of a formal uncertainty budget. Comparison against an absolute frequency reference rather than another spectroscopic calibration method subject to scan nonlinearity and drift errors is also required to provide an absolute measure of the method's accuracy. Future work may also entail the mitigation of these yet unquantified error sources in order to provide (1) a more accurate calibration and (2) a better understanding of environmental effects and/or artifacts of the experimental setup which contribute to shifts and/or broadening in spectral features. Related to point (2), another plausible avenue for future work is a full treatment of an ensemble of multi-level atoms via the density matrix formalism to investigate how changes in parameters such as coupling strengths or temperature, or the incorporation of additional structure such as magnetic sub-levels, may impact the efficacy of the two-level approximation under various operating conditions.

\bigskip
\section*{Acknowledgments}
This research was developed with funding from NIST. The views and opinions of authors expressed herein do not necessarily state or reflect those of the United States Government or any agency thereof. A contribution of the U.S. government, this work is not subject to copyright in the United States. 

\section*{Conflict of Interest}
The authors declare no conflicts of interest.

\section*{Data Availability Statement} 
All of the data presented in this paper is available at https://doi.org/XX.XXXXX/XXXX-XXXX

\bibliography{bibliography}

\end{document}